\documentclass[a4paper,11pt]{article}

\usepackage{amssymb,amsmath}

\usepackage{graphicx}
\usepackage{subfig}

\usepackage[left=1in,right=1in,top=1in,bottom=1in]{geometry}

\usepackage[dvipsnames]{xcolor}

\usepackage{hyperref}
\hypersetup{
  colorlinks   = true, 
  urlcolor     = OliveGreen, 
  linkcolor    = red, 
  citecolor    = blue 
}

\title{Power asymmetry in CMB polarization maps from PLANCK : a local variance analysis}

\author{Pavan K. Aluri$^{1,2}$\footnote{aluri@kasi.re.kr}, Arman Shafieloo$^{2,3}$\footnote{shafieloo@kasi.re.kr }}

\date{\today}

\begin{document}

\maketitle

\centerline{$^1$Korea Institute for Advanced Study (KIAS), Quantum Universe Center (QUC),}
\centerline{85 Hoegi-ro, Dongdaemun-gu, Seoul - 02455, Republic of Korea}
\centerline{$^2$Korea   Astronomy   and   Space   Science   Institute (KASI),}
\centerline{Yuseong-gu, 776 Daedeok daero, Daejeon 34055, Republic of Korea}
\centerline{$^3$University of  Science and Technology (UST),}
\centerline{Yuseong-gu 217 Gajeong-ro, Daejeon 34113, Republic of Korea}
 
\begin{abstract}
A persistent signal of power asymmetry on opposite hemispheres of CMB sky
was seen in full-sky temperature measurements made so far. This asymmetry
was seen in microwave sky from WMAP as well as PLANCK satellites, and calls for attention the larger question of \emph{statistical isotropy}, one of the
foundational principles of modern cosmology.
In this work we present an analysis of polarized CMB maps from PLANCK 2015
full mission data. We apply the local variance estimator on low resolution
$E-$mode maps from PLANCK 2015 polarization \texttt{Commander} solution.
We find a significant hemispherical power asymmetry in polarization
data on large angular scales, at the level of $\sim 2.6-3.9\%$ depending
on the galactic mask, and the circular disc radius used for computing local
variance maps. However the direction is found to be pointing
broadly towards CMB kinetic dipole direction. Precise measurements of CMB
polarization in future will shed light on this apparent discrepancy in the
anisotropy axis seen in temperature and polarized CMB sky, and likely influence
of systematics on our findings.
\end{abstract}

\section{Introduction}
Hemispherical power asymmetry (HPA), initially observed in WMAP first year temperature
data~\cite{eriksen04} has come to be one of the outstanding anomalies that indicated
violation of \emph{statistical isotropy} on large angular scales of CMB sky.
This anisotropic signal persisted between completely different full-sky missions
viz., WMAP and PLANCK probes, that involved different analysis procedures,
systematics and science teams~\cite{wmap7anom,plk13is,plk15is}.
HPA attracted significant attention from the cosmology community, which was estimated
using a variety of methods, see for example Ref.~\cite{hansen04, hajian05, prunet05, picon06,
eriksen07, bernui07, lew08, bernui08, hansen09, hanson09, paci10, paci13, flender13,
rath13, santos14, akrami14, bernui14, adhikari15, quartin15, aiola15, ghosh16}.
Many explanations were also put forth to explain this anomaly with no final
conclusion.

This hemispherical power asymmetry has come to be modelled as a dipole
modulation of otherwise statistically isotropic CMB sky as \cite{gordon05,gordon07},
\begin{equation}
\Delta\tilde{T}(\hat{n})= \Delta T (\hat{n})(1+A\,\hat{\lambda}\cdot \hat{n})\,,
\label{eq:mod-temp-aniso}
\end{equation}
where $\Delta\tilde{T}(\hat{n})$ and $\Delta T(\hat{n})$ are the modulated/observed
and statistically isotropic CMB anisotropy fields respectively in the direction $\hat{n}$.
$A$ and $\hat{\lambda}$ are the amplitude and direction of modulation.
Here the amplitude can be scale dependent~\cite{plk13is,plk15is,rath13,akrami14,aiola15,ghosh16}.
This phenomenological modelling of HPA as dipole modulation of temperature
field has important implications for the statistics of observed CMB signal.
It induces off-diagonal correlations in the two-point function that couples a
multipole $l$ to $l'=l \pm 1$. These correlations were exploited to estimate
the anisotropic signal itself~\cite{plk13is,plk15is,prunet05,hanson09,rath13,aiola15,quartin15,ghosh16}.

In this paper, we probe the presence of hemispherical power asymmetry in
the polarized CMB signal from PLANCK full mission data. The PLANCK
science team didn't report any analysis of PLANCK 2015 CMB polarization maps
in the context of isotropy studies. As the PLANCK papers inform~\cite{plk15cmb},
this is because of the uncertainty in the level of recovered polarization
signal on large angular scales, that warrants such analysis. Accordingly the
low multipoles of polarized CMB signal in the various component separated CMB
polarization maps have been excluded in the full mission PLANCK data that is made public.
Further the complimentary simulations to polarization data that are also made public,
have a significant level of mismatch in the noise level compared to data.
However in this paper, we work in the lower to intermediate range of multipoles,
and suitably modifying the noise simulations, that are already made available through
the second public release to compute significances.

\section{CMB polarization maps}

The CMB temperature anisotropies are conventionally expanded in terms of spherical
harmonics as
\begin{equation}
\Delta T (\hat{n}) = \sum_{l=2}^\infty \sum_{m=-l}^{l} a^T_{lm} Y_{lm} (\hat{n})\,,
\end{equation}
where $\Delta T (\hat{n})$ are temperature anisotropies after subtracting the
leading monopole and dipole components in the sky direction $\hat{n}$, $a^T_{lm}$
are the coefficients of expansion, and $Y_{lm}(\hat{n})$ are a suitable basis
to expand the temperature anisotropies on a sphere.

Unlike the intensity field, CMB polarization stokes $Q/U$ parameters are not scalars
under rotation and are frame dependent. The combination $X_\pm (\hat{n}) = (Q \pm iU) (\hat{n})$
transforms as $X'_\pm (\hat{n}) = e^{\mp 2i\psi}(Q \pm iU) (\hat{n})$, under rotation
by an angle $\psi$ of the co-ordinate plane  perpendicular to the direction $\hat{n}$
in which $Q/U$ are defined. Hence $X_\pm (\hat{n})$ transform as a spin~$\pm 2$ field
under coordinate transformations. Thus it is expanded in terms of spin-weighted spherical harmonics,
${_s}Y_{lm}$, as
\begin{equation}
X_\pm (\hat{n}) = (Q \pm i U) (\hat{n})
 = \sum_{l=2}^\infty \sum_{m=-l}^{l} a_{\pm 2,lm} \,\, _{\pm 2}Y_{lm} \hat{n}\,,
\end{equation}
where $_{\pm 2}Y_{lm} (\hat{n})$ are the spin-2 spherical harmonics, and $a_{\pm 2,lm}$
are the coefficients of expansion in that basis~\cite{gb67,zs97,kks97,z98}.
$Y_{lm} = {_0}Y_{lm}$ are the usual spherical harmonics that are used to decompose
the CMB temperature field which is a spin-0 field.

Further, a linear combination of these spin-2 spherical harmonic coefficients, $a_{\pm 2,lm}$,
can be defined as~\cite{zs97,z98}
\begin{eqnarray}
a^E_{lm} &=& -(a_{2,lm} +a_{-2,lm})/2 \quad \textnormal{and} \\
a^B_{lm} &=& -(a_{2,lm} +a_{-2,lm})/2i\,,
\end{eqnarray}
where $a^E_{lm}$ and $a^B_{lm}$ are respectively called \emph{electric} and \emph{magnetic}
modes of polarization that are invariant under
rotations. Similar to an electric and magnetic field, $a^E_{lm}$ and $a^B_{lm}$ respectively
have even and odd parity properties under inversion ie., for $\hat{n}\rightarrow -\hat{n}$,
$a^E_{lm} \rightarrow (-1)^la^E_{lm}$ and $a^B_{lm} \rightarrow (-1)^{l+1}a^B_{lm}$.
Thus one can obtain two spin-0 fields as
\begin{equation}
E(\hat{n}) = \sum_{lm} a^E_{lm} Y_{lm} (\hat{n}) \quad \textnormal{and} \quad
B(\hat{n}) = \sum_{lm} a^B_{lm} Y_{lm} (\hat{n})\,.
\label{eq:ebmaps}
\end{equation}
As mentioned above, owing to their parity properties under inversion,  $E(\hat{n})$ is a scalar field
while $B(\hat{n})$ is a pseudo-scalar field. Physically these modes originate from distinct
sources. Whereas, scalar perturbations give rise to purely $E-$modes, only tensor
perturbations can generate $B-$modes (on large angular scales).
Lensing due to large scale structure can induce $B-$modes from $E-$modes at small angular scales.

In this paper, we test for hemispherical power asymmetry in CMB $E-$mode maps derived from
PLANCK full mission data.

\section{Methodology and Data sets}
\subsection{Local variance estimator}\label{sec:lve}
We use the recently proposed local variance estimator (LVE) that was applied to probe hemispherical power asymmetry in temperature data~\cite{akrami14}. This statistical approach is based on the Crossing statistic that has been proposed to address some different problems in astronomy \cite{Crossing1,Crossing2,Crossing3,Crossing4}.

Hereafter we use LV or LVE to refer to local variance analysis or maps thus estimated.
In this section, we briefly describe the LV method to apply on polarized CMB maps.

One can define a variance map of the observed CMB temperature sky as,
\begin{equation}
\sigma^2_{r} (\hat{N}) = \sum_{p\,\in \, r@\hat{N}} (T(p) - \bar{T}_r)^2\,,
\label{eq:lvestimator}
\end{equation}
where $\sigma^2_{r}$ is the variance computed locally from all the
pixels $p$, falling inside a circular disc of chosen radius $r$ defined
in the sky direction $\hat{N}$, and $\bar{T}_r$ is the mean temperature
corresponding to the same circular region.

A variance map of dipole modulated CMB sky following Eq.~[\ref{eq:mod-temp-aniso}]
is given by
\begin{equation}
\tilde{\sigma}^2 (\hat{N}) \approx \sigma^2 (\hat{N}) (1+2\,A\,\hat{\lambda}\cdot \hat{N})\,,
\label{eq:lve-dip-mod}
\end{equation}
upto first order in the isotropy violating dipole field, where $\tilde{\sigma}^2$ and $\sigma^2$
are maps of locally computed variances of modulated and unmodulated CMB anisotropies respectively.
Further, a normalized variance map is constructed as
\begin{equation}
\xi (\hat{N}) = \frac{\sigma^2_{obs} (\hat{N}) - \langle \sigma^2 (\hat{N}) \rangle}{\langle \sigma^2 (\hat{N}) \rangle}\,,
\label{eq:normlvemap}
\end{equation}
where $\langle \sigma^2 (\hat{N}) \rangle$ is the expected mean field bias in variance estimated
from CMB simulations (including inhomogeneous noise), that is subtracted from the local variance
map of the observed sky.
From this normalized variance map we compute the dipole amplitude (and direction) to check
for consistency with random expectations for any given realization.
Any significant dipole amplitude thus found in data would indicate a significant hemispherical
power asymmetry, hemisphere defined by the corresponding dipole direction, and hence a
breakdown of statistical isotropy. We should note that one can define different metrics to test the statistical isotropy using the normalized variance map but in this work we focus on the dipole amplitude and its direction.   

In practice, one implements the local variance estimator as follows. Any cleaning procedure
to clean the raw satellite data, is limited by the level of foreground emission in a given
sky region. Correspondingly, owing to high emission levels in the galactic plane, the recovered
CMB signal is less reliable in those regions. Bright point and extended sources of
extragalactic origin also lead to the same situation. Thus these regions are masked in any
CMB analysis to avoid biases due to residual foreground contamination. Hence, in computing
the local variances at different sky locations, the circular disc mask defined locally
is taken in union with the galactic mask to exclude potentially contaminated regions
of the recovered CMB sky. Further to make the analysis robust, we can define a criterion
for using these locations only if a chosen fraction of the sky pixels, say for example 50\% or
80\%, in comparison to the number of pixels available using the full disc mask, survive in
the effective disc mask (ie., the union of galactic mask and the locally defined circular
disc mask of a chosen radius). Thus, the fraction of the sky over which local variances
are computed is less than the sky fraction of the galactic mask used to omit regions
with potential residual contamination. Operationally, we set these invalid pixels to the \texttt{HEALPix}
bad value $=-1.6375 \times 10^{30}$.  And a dipole is fit to the (normalized) local variance
map thus obtained using only the valid pixels.

To analyze the polarization maps, we apply the LV estimator to the
$E-$mode map (defined by Eq.~[\ref{eq:ebmaps}]). The variance is computed as
$\sigma^2_r (\hat{N}) = \langle E^2(p) \rangle_{p\in r @\hat{N}}$, where the angular
brackets denote the average of the squared $E-$mode field computed using all pixels, $p$,
falling inside an effective circular disc of chosen radius, $r$, defined at $\hat{N}$
on the CMB sky.

\subsection{Data sets used}
\subsubsection{$E-$mode CMB maps}
\label{sec:cmb-obs-map}
We do our local variance analysis on a CMB $E-$mode map. To this extent
we use the full-sky Stokes $Q$ and $U$ CMB maps from PLANCK full mission data, estimated
using the \texttt{Commander} algorithm. These $Q/U$ maps are made available through PLANCK's
second public release\footnote{\url{http://irsa.ipac.caltech.edu/data/Planck/release_2/all-sky-maps/matrix_cmb.html}}.
Using partial skies to go from $Q/U$ to $E/B$ basis, will induce undesired correlations
due to masking in the resultant maps. Hence we use full-sky polarization maps to obtain $E-$map.
PLANCK also employed other component separation schemes viz., \texttt{NILC}, \texttt{SEVEM},
and \texttt{SMICA}~\cite{plk15cmb} to extract CMB signal from raw satellite data.
But the recovered Stokes $Q/U$ CMB maps from these other methods have a portion of the sky removed,
or some visible contamination still present, in the galactic region.  Hence we use only the
\texttt{Commander} Stokes $Q$ and $U$ maps to get full-sky $E-$mode map in this work. Note that we
eventually use galactic mask in our studies which excise unreliable regions of the recovered CMB sky
from raw data using PLANCK provided polarization mask. We also make use of custom-made
polarization masks, as explained later.

The \texttt{Commander} estimated $Q/U$ maps are available at \texttt{HEALPix} $N_{side}=1024$
with smoothing given by a Gaussian beam of $FWHM=10'$ (arcmin).
Since the polarized CMB maps are dominated by noise are high$-l$, in this paper we work with
a low resolution $E-$mode map at $N_{side}=256$ with a beam resolution of $FWHM=40'$ (arcmin)
Gaussian beam.
First the \texttt{HEALPix} facilities - \texttt{anafast} and \texttt{alteralm} - are applied in
succession to obtain $E/B$ $a_{lm}$'s from Stokes $Q/U$ maps, at the desired lower beam/pixel
resolution.

Now, as mentioned earlier, various component separated maps have their low$-l$
modes removed in the maps that are made publicly available. This is so because of the
unsatisfactory level of systematic artefacts still present on large angular scales of
PLANCK polarization maps~\cite{plk15cmb}. A window function in multipole space defined as
\begin{equation}
f_l = \left\{ \begin{array}{l l}
                 0 & \forall \quad l<l_1\\
                 \frac{1}{2}\left( 1-\cos\left(\pi \frac{l-l_1}{l_2-l_1} \right) \right] & \forall \quad l_1 \leq l \leq l_2\\
                 1 & \forall \quad l>l_2
               \end{array} \right.
\label{eq:highpassfilter}
\end{equation}
where $l_1=20$ and $l_2=40$, is applied to these publicly released CMB maps.
Further as discussed in Appendix~\ref{apdx:noisesim}, to
reduce the effect of noise dominated modes in the CMB $E-$mode map thus
obtained on our analysis, we additionally apply a similar cosine filter at high$-l$ as
\begin{equation}
f_l = \left\{ \begin{array}{l l}
                 1 & \forall \quad l<l_1\\
                 \frac{1}{2}\left[ 1+\cos\left(\pi \frac{l-l_1}{l_2-l_1} \right) \right] & \forall \quad l_1 \leq l \leq l_2\\
                 0 & \forall \quad l>l_2
               \end{array} \right.
\label{eq:lowpassfilter}
\end{equation}
where $l_1=220$ and $l_2=240$.

\begin{figure}
\centering
\includegraphics[width=0.9\textwidth]{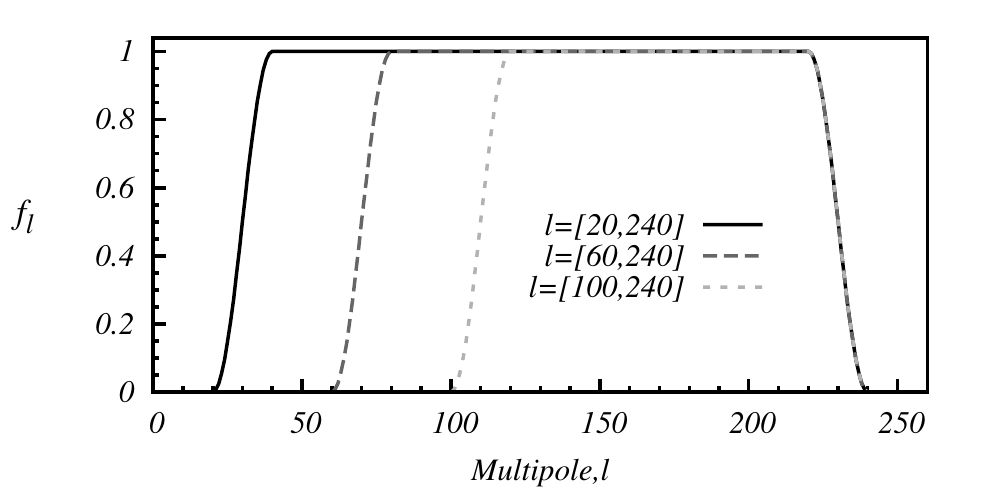}
\caption{Three different high-pass filters used to produce $E-$mode CMB maps, all
         with a fixed low-pass filter are shown here. These different filters are applied
         to produce polarized $E-$maps comprising different angular scales to probe for any
         scale dependence in the amplitude of dipole, fit to the derived local variance
         maps. At high$-l$, the low-pass filter is applied to smoothly suppress the effect of
         noisy modes in data. See text for details.}
\label{fig:lfilters}
\end{figure}

The hemispherical power asymmetry, as found in CMB temperature data is present
significantly on the large angular scales viz., $l \lesssim 64$~\cite{plk13is,plk15is,
hanson09,rath13,aiola15,quartin15,ghosh16}.
In order to test for similar behaviour in the polarized CMB map, we do additional high pass filtering
to get two $E-$mode CMB maps using the window function defined by Eq.~[\ref{eq:highpassfilter}] with $l_1=60$/$l_2=80$ and $l_1=100$/$l_2=120$ at low$-l$. The various $l-$space windows used to produce
the filtered $E-$mode CMB maps are shown in Fig.~[\ref{fig:lfilters}].

These filtered $E-$mode $a_{lm}$'s are then converted to an $E-$mode map as
\begin{equation}
E(\hat{n}) = \sum_{lm} f_l \, a^E_{lm} \, Y_{lm} (\hat{n})
\label{eq:filteredEmap}
\end{equation}
using \texttt{HEALPix} facility \texttt{synfast}.
Thus, in the current study, we use three different high pass filtered
CMB $E-$mode maps as derived from PLANCK provided
Stokes $Q/U$ maps estimated using \texttt{Commander}
algorithm. They have a beam resolution given by a Gaussian beam of $FWHM=40'$ (arcmin) and
at $N_{side}=256$, comprising different angular modes.

\subsubsection{Masks}\label{sec:pmasks}
Common galactic masks for polarization analysis were provided by the PLANCK team through
their second data release\footnote{\url{http://irsa.ipac.caltech.edu/data/Planck/release_2/ancillary-data/}} at \texttt{HEALPix} $N_{side}=1024$. The confidence masks provided by the individual component separation
methods were combined to obtain a union mask which would be the most conservative mask
for CMB analysis irrespective of the specific component separation method used to extract
CMB signal. This union mask (referred as UP78) was further extended to obtain two versions UPA77 and UPB77.
We use the UPB77 mask in the current work, which was deemed more reliable than the other
in handling point source contamination. This mask UPB77 has a sky fraction
of $\approx 77\%$. See Ref.~\cite{plk15cmb} for more details regarding PLANCK galactic masks.

\begin{figure}
\centering
\includegraphics[width=0.75\textwidth]{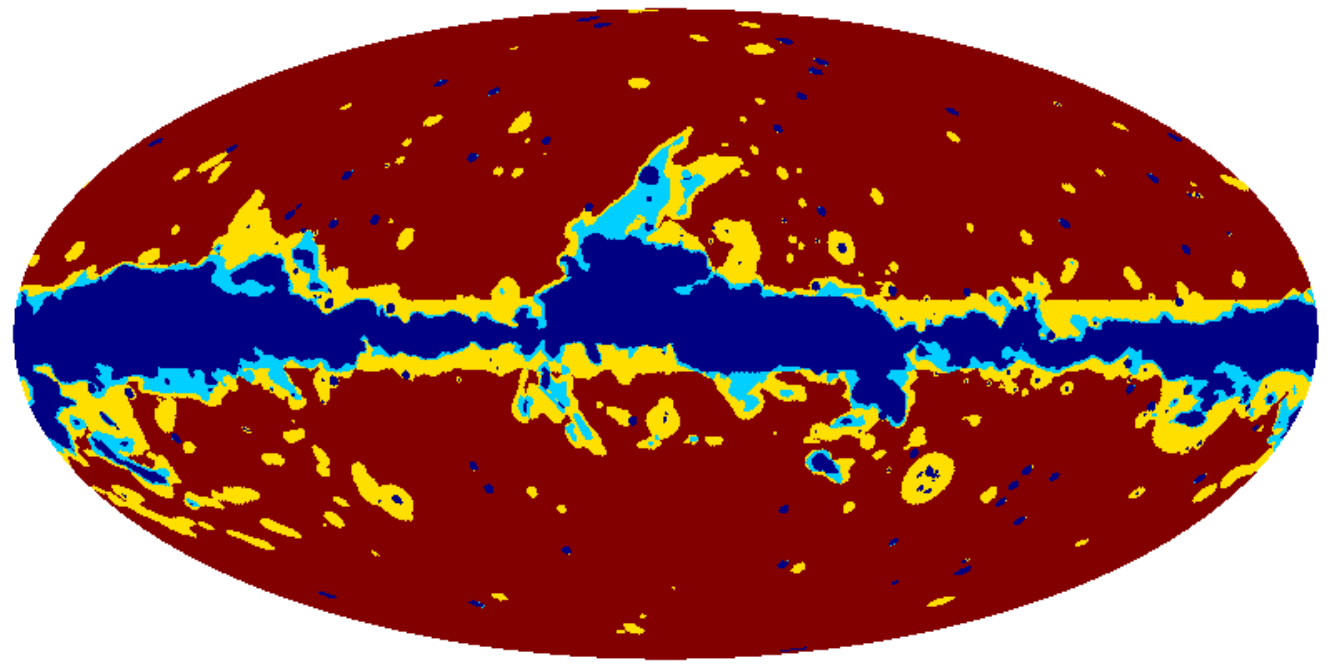}
\caption{Various polarization masks used in the present work are shown here.
         These are created in $P$-space where $P^2=\sqrt{Q^2 +U^2}$. The \emph{deep blue}
         region is excluded by all masks. The \emph{red} region corresponds to a mask
         derived by downgrading PLANCK's UPB77 mask, which has an $f_{sky} \approx 74\%$.
         The \emph{light blue} and
         \emph{yellow} regions (including \emph{red}) are our custom made polarization masks which
         have an available sky of $\approx 80\%$ and $85\%$ respectively. These three masks are
         referred to as \texttt{mask074}, \texttt{mask080}, and \texttt{mask085} respectively.
         Point sources are also accounted for in these masks.}
\label{fig:masks}
\end{figure}

Since we work at low resolution of \texttt{HEALPix} $N_{side}=256$, the UPB77 mask is
downgraded as follows.
The high resolution mask is available
at \texttt{HEALPix} $N_{side}=1024$. We first invert this mask and fill all island like
regions inside using the \texttt{HEALPix} facility \verb+process_mask+.
To be more conservative, these island like regions in the galactic region are excluded to begin with.
This inverted/filled high resolution mask is downgraded
to $N_{side}=256$. Then a Gaussian beam of $FWHM=40'$ (arcmin) is applied, which is
same as that of the CMB map we analyse. Finally a cut-off of 0.05 is applied on this inverted,
filled, downgraded, and smoothed mask where those pixels with pixel-value less than this cut-off
are set to `1', and `0' otherwise.
 We do so, so as to be able
to handle leakage due to smearing of extended/point source contamination due to smoothing.
The mask thus obtained has a sky fraction of $f_{sky}\approx 0.74$.
This mask is referred to as \texttt{mask074}, and is shown in \emph{red} in Fig.~[\ref{fig:masks}].

Now the confidence masks for individual component separated CMB polarization maps
have an available sky fraction of $\approx 83\%$, $96\%$, $79\%$ and $85\%$ respectively
for \texttt{Commander}, \texttt{NILC}, \texttt{SEVEM}, and \texttt{SMICA} derived solutions.
Their mask contours are different, and hence they excise different regions of the sky.
So, we further devise two more masks with larger sky fraction for use in this analysis as follows.
Here we adopt the procedure of Tegmark et al. (2003)~\cite{tegmark03}, used to define sky
partitions for iterative cleaning of the CMB sky.
To create these additional masks, we use the raw polarization maps of PLANCK \texttt{030}, \texttt{070}
and \texttt{353} GHz channels. These maps with different beam/pixel smoothing and map resolution
are first downgraded to have a common beam, pixel and map resolution. The raw $Q/U$ maps
corresponding to each of these three frequencies are deconvolved with their respective
circular beam transfer functions,
and simultaneously smoothed with a Gaussian beam of $FWHM=2^\circ$ (degrees),
so as to have largely contiguous regions in the masks being created,
while repixelizing them at a lower resolution of $N_{side}=256$.
Among the PLANCK frequency channels, the \texttt{070} GHz band has visibly lowest
foreground levels in the raw satellite data. The raw map corresponding to this frequency
band is subtracted from the lowest and highest frequency raw maps from PLANCK that are
sensitive to polarization viz., the \texttt{030} and \texttt{353} GHz channels. These two
difference maps, \texttt{030-070} and \texttt{353-070}, in $Q/U$ basis
containing only foregrounds,
are dominated by synchrotron and thermal dust respectively.
The individual difference maps in $Q/U$ are further combined to form $P-$maps, where $P^2 = Q^2+U^2$.
These two $P-$maps, are then used to create what is called
a \emph{junk map} which retains only the pixel-value that is largest between
the two $P-$maps at any sky location.
This junk map taken as representative of the maximum foreground signal in any
sky direction due to any polarized non-CMB emission is thresholded at different
values to obtain galactic masks with different sky fractions.
Thus a suitable cut-off is applied to this junk map to obtain two more
galactic masks that have a sky fraction of $80\%$ and $85\%$, respectively.

In order to account for the point source contamination, we extend the point source masks
in polarization as provided by PLANCK team at different frequencies\footnote{\url{http://irsa.ipac.caltech.edu/data/Planck/release_2/ancillary-data/}}.
The point source masks that are available at $N_{side}=2048$ at different frequency bands
are first added together to make a composite polarized point source mask. This composite mask
at higher $N_{side}$ is then processed similar to the galactic masks described above to obtain
an  extended composite point source mask at low resolution of $N_{side}=256$ that is then
applied to the two galactic masks we created viz., \texttt{mask080} and \texttt{mask085}.
Although the addition of point sources reduces the sky fraction,
we continue to refer to them as \texttt{mask080} and \texttt{mask085}.
The \emph{yellow} and \emph{light blue} regions (along with \emph{red} region) shown in  Fig.~[\ref{fig:masks}], correspond to these two masks respectively.

Thus we have three masks labelled \texttt{mask074}, \texttt{mask080}, and \texttt{mask085}
which have respectively $f_{sky} \approx 74\%$, $80\%$, and $85\%$
unmasked sky fractions. The LVE maps are obtained using these three foreground exclusion masks.
We additionally created two more masks in $E-$basis, and the whole LV analysis is repeated using these
$E-$masks as test of robustness of our analysis using $P-$masks.

\subsubsection{Simulations}\label{sec:sim}
Here we describe the generation of  simulations used to supplement the
clean polarized CMB maps from satellite observations for computing significances.

PLANCK team provided a set of 1000 CMB maps at each frequency band in which
it made observations of polarized microwave sky, that have same instrument properties
such as beam smoothing similar
to the observed maps. Also provided are a set of 1000 inhomogeneous noise maps corresponding
to each of the seven frequency bands (\texttt{030} - \texttt{353} GHz). These are called Full Focal
Plane (FFP) simulations.
However the PLANCK collaboration didn't make available, a similar set of processed CMB
maps with noise, that complement polarized CMB sky recovered from data using various
component separation methods.
Hence we process the high resolution CMB and noise FFP simulations provided, to suitably
obtain mock maps that augment the observed $E-$mode CMB maps used in the present
study.

The frequency specific Stokes $Q/U$ CMB and noise maps are available at a \texttt{HEALPix}
resolution of $N_{side}=1024$ or $2048$, and have different beam smoothing, depending on the
frequency channel. Hence all these maps are brought to a lower resolution of $N_{side}=256$,
and to have beam resolution given by a Gaussian beam with $FWHM=40'$ (arcmin) following the procedure
outlined in Sec.~[\ref{sec:cmb-obs-map}] (ie., applying \texttt{HEALPix} facilities \texttt{anafast}
and \texttt{alteralm} in sequence).
In this process, we used the polarization circular beam transfer functions ($b_l$) for deconvolution, that are also provided with PLANCK's second public release\footnote{\url{http://irsa.ipac.caltech.edu/data/Planck/release_2/ancillary-data/}}.
The polarized spherical harmonic
coefficients, $a^E_{lm}$, of a simulated CMB map thus obtained are converted to an $E-$mode map using the
\texttt{HEALPix} facility \texttt{synfast}.

The polarization (inhomogeneous) noise maps are also
subjected to the same procedure to obtain low resolution $E-$mode noise maps. Now, the
smoothed/downgraded $E-$mode noise maps are combined to obtain noise r.m.s
maps corresponding to each channel at $N_{side}=256$.
These  noise r.m.s maps are then used to obtain a set of 1000 mock maps
following inverse noise variance addition of $E-$mode CMB and noise maps corresponding to different
frequency channels with same simulation/seed number\footnote{In the absence of foregrounds,
an ILC procedure will reduce to an inverse noise variance
weighing if the pixels are uncorrelated i.e., $E_{co-add}(p) = \sum_{i=1}^{N_c} w_i(p) E_i(p) / \sum_{i=1}^{N_c} w_i(p) $,
where $i$ denotes the $N_c(=7)$ frequency channels, $w_i(p)=1/\sigma^2_i(p)$ are the weights
which are given by the inverse of r.m.s squared of smoothed/downgraded polarized noise maps at
$N_{side}=256$ at each frequency, and $p$ denoted a pixel/sky direction in the sky.
Hence ignoring the correlations between pixels due to beam,
we  used the diagonal approximation to the noise covariance matrix to obtain suitable simulations
that complement the observed maps. This is justified following our interest in probing large angular
scales of the polarized CMB sky. However the effective polarization noise maps thus obtained
from FFP simulations are still not usable due to an additional complication. See text for more
details.}.

However the effective noise maps thus obtained are not yet useful due to an underestimation
of noise in FFP simulations compared to data. As described in Appendix~\ref{apdx:noisesim},
we do a rescaling of the effective noise maps by a constant multiplicative factor $\approx 1.32$.
Thus a set of 1000 mock CMB $E-$mode maps with noise are generated complementing the observational
data. These co-added low resolution $E-$mode CMB maps with (scaled) noise are further filtered
using the high and low pass cosine filters defined in Eq.~[\ref{eq:highpassfilter}] and [\ref{eq:lowpassfilter}].
Thus from each co-added \emph{realistic} $E-$mode CMB realization, we obtain three filtered maps
similar to the observed data maps described in Sec.~[\ref{sec:cmb-obs-map}].

\section{Results}\label{sec:results}
In the present analysis, we use a CMB
$E-$mode map derived from PLANCK 2015 Stokes $Q/U$ CMB maps, estimated using \texttt{Commander}
algorithm. Different window functions in multipole space are applied to this $E-$mode CMB map
to get three filtered maps, as described in Sec.~[\ref{sec:cmb-obs-map}]. In order
to probe any scale dependent behaviour of the amplitude of hemispherical power
asymmetry, if present, that is modelled as dipole modulation of otherwise statistically
isotropic CMB anisotropies, we generated these three filtered maps.
All the three maps have a low-pass cosine filter applied with $l_1/l_2=220/240$ following
Eq.~[\ref{eq:lowpassfilter}]. A high-pass cosine filter defined by Eq.~[\ref{eq:highpassfilter}]
with $l_1/l_2=20/40$, $60/80$ and $100/120$ are applied to obtain the said three filtered $E-$mode CMB maps.
Note that the $l_1/l_2=20/40$ filter is already applied by PLANCK team to the publicly released
polarization CMB maps. We only generated the later two filtered CMB $E-$maps.
As a consistency check of our results, we also performed the LV analysis on large angular
scales using foreground masks derived in $E-$space directly.

\subsection{Power asymmetry on large angular scales}\label{subsec:lvelowl}

Here we present results from analyzing the CMB maps containing large angular scales ie., using
the filtered $E-$mode CMB map comprising the multipoles $l=[20,240]$. We used four
different circular disc of radii $120'$, $90'$, $60'$ and $45'$ (arcmin), to compute
local variances. The largest angular mode that is fully available in the polarized CMB map
from the second data release is $l=40$, and corresponds to an angular size of $\theta = 180^\circ/l = 180^\circ/40 = 4.5^\circ$ (degrees). So we limit to using a maximum disc radius of $2^\circ$ (degrees)
ie., a maximum circular disc of diameter $4^\circ$ (degrees).
The circular discs of different radii are defined at the pixel centers of a \texttt{HEALPix}
grid of $N_{side}=64$, to compute the local variances. An $N_{side}=64$ map has pixels of size $a=\sqrt{4\pi/(12\times 64^2)}\approx 55'$ (arcmin) assuming them to be a square of side $a$.
Later on we also use two more filtered $E-$mode CMB maps, and given the angular scales they contain,
we compute all the local variance maps using different disc radii at the same \texttt{HEAPix}
resolution of $N_{side}=64$.
In practice the local variances are computed from effective circular discs meaning the circular discs
defined locally are taken in conjugation with a galactic mask to avoid potential foreground
bias on our analysis. We compute local variances only if the effective discs have atleast
$90\%$ of the pixels in comparison to those in full disc so that our statistics are robust.
This results in local variance maps with less sky fraction that the galactic mask itself used to
exclude regions with potential galactic contamination in the data map. The three $P-$masks
shown in Fig.~[\ref{fig:masks}] are used in the present study.

The filtered \texttt{Commander} 2015 $E-$mode CMB map comprising the large angular scales
as available, is used to compute four local variance maps at $N_{side}=64$, following Eq.~[\ref{eq:lvestimator}], corresponding to the choice of four circular disc of radii
$120'$, $90'$, $60'$ and $45'$ (arcmin) described earlier.
These LV maps are further corrected for and normalized using \emph{mean field bias} computed
from simulations. Thus we obtain normalized local variance maps as defined by Eq.~[\ref{eq:normlvemap}].
A dipole is then fit to these normalized local variance maps from data, assuming that the
power asymmetry, if any, is dipolar ($L=1$) in nature. We eventually check for presence of any
significant power in higher modes ($L>1$) of local variance maps, that could be inducing power
asymmetry.

In order to compute the \emph{mean field bias} to correct the local variance maps from data,
we proceed as follows. The generation of simulations with same beam smoothing and consistent
noise levels as data, are described in Section~[\ref{sec:sim}] and Appendix~[\ref{apdx:noisesim}].
The simulations thus obtained are filtered to contain the same multipole range as data by applying
the high and low pass cosine filters following Eq.~[\ref{eq:highpassfilter}],
[\ref{eq:lowpassfilter}] and [\ref{eq:filteredEmap}].

We then apply the local variance estimator to simulations similar to data ie., computing
local variance maps at \texttt{HEALPix} $N_{side}=64$ with the same choice of circular disc radii
and effective disc fraction of $90\%$ described above, from each simulation.
Thus a set of 1000 local variance maps are derived,
corresponding to each of the circular disc radii chosen and the galactic mask used. A \emph{mean field}
is thus estimated by taking the mean of the 1000 LV maps from simulations for each choice of disc radius
and mask. This is subtracted from and also used to normalize the data LV maps (Eq.~[\ref{eq:filteredEmap}]).
The simulations themselves are also normalized using the mean field LV estimate.

\begin{figure}
\centering
\includegraphics[width=0.88\textwidth]{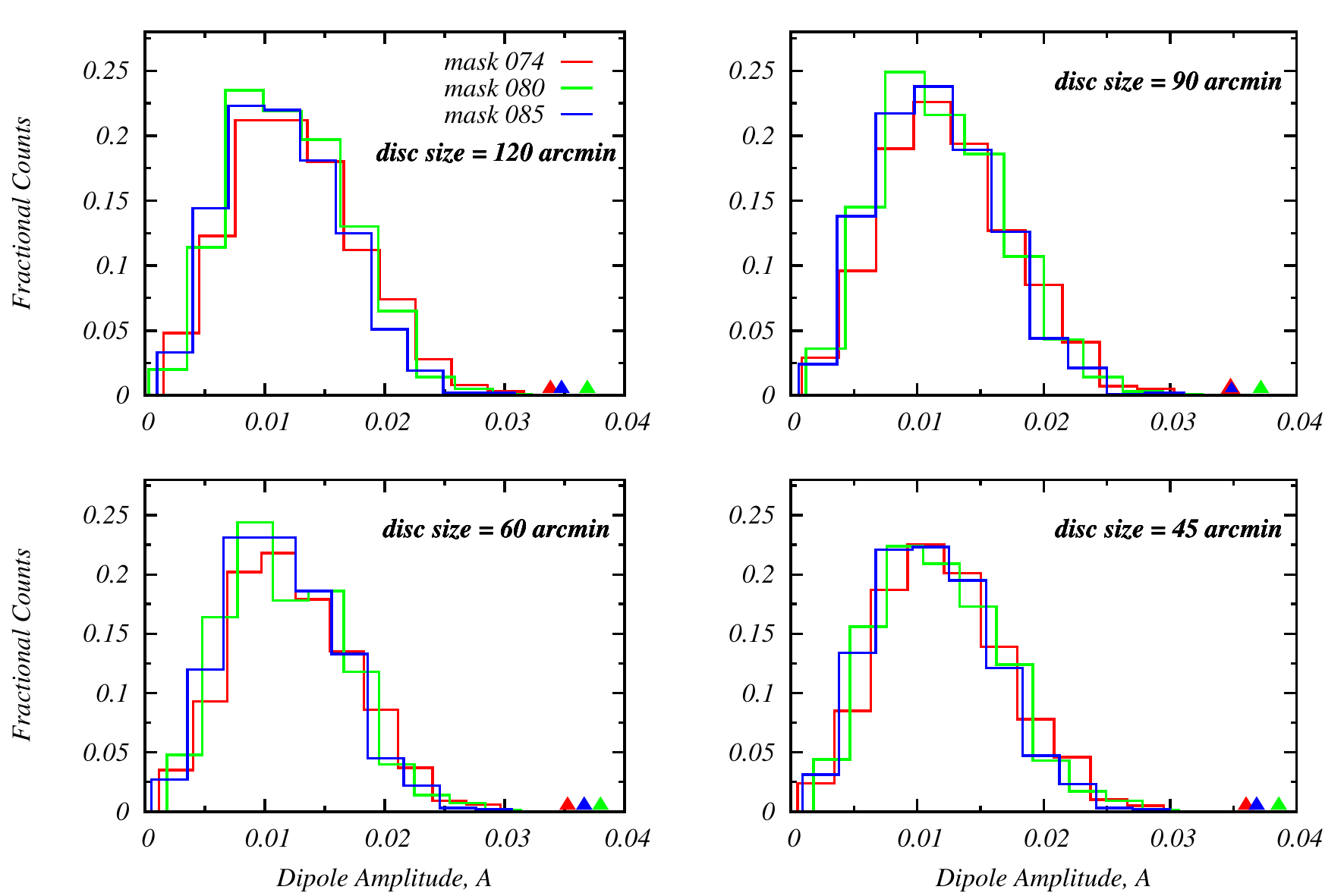}
\caption{Histograms of amplitude of dipole, fit to the normalized local variance maps from
         simulated $E-$mode CMB maps are shown here, in comparison to data
         values (denoted by small triangles in each plot). The $E-$mode polarization map
         comprising the multipole range $l=[20,240]$ is analyzed here using four circular discs
         of radii $120'$. $90'$ $60'$ and $45'$ (arcmin), defined locally in each sky direction
         for computing local variances.
         Each plot has three histograms that correspond to the three $P-$masks with different
         sky fractions, used to estimate the LV maps.}
\label{fig:dip-mod-ampl}
\end{figure}

The amplitudes of dipole power asymmetry in data in comparison to simulations are shown in
Fig.~[\ref{fig:dip-mod-ampl}]. The four plots in that figure correspond to the four choices
of disc radii made to compute the local variances across the sky.
The \emph{red}, \emph{green} and \emph{blue} histograms in each plot of the figure correspond
to the three $P-$masks described in the previous section (Fig.~[\ref{fig:masks}]).
The amplitude of dipole in LV maps from data are denoted by triangle point types in
each plot. We used the \texttt{HEALPix}
subroutine \verb+remove_dipole+ to obtain the dipole amplitudes and directions from data
and simulations. An estimate of the variance of the normalized local variance maps from
isotropic simulations is used for inverse noise variance weighing of pixels to compute
the dipole amplitudes and directions from normalized LV maps while using the subroutine
\verb+remove_dipole+.
The dipole amplitudes as well as probability-to-exceed the observed values ($p-$values)
corresponding to different choices of circular disc radii for mapping the local variances
in \texttt{Commander} 2015 CMB $E-$mode map comprising large angular scales ($l=[20,240]$)
are tabulated in the top block of Table~\ref{tab:dip-ampl-pval}. The three rows in that
block correspond to the three $P-$masks used. The dispersion in recovered dipole amplitudes
from corresponding simulations are quoted as errors in braces in the table.

\begin{table}
\centering
\begin{tabular}{c c c c c}
\hline
$l-$range & ds=$120'$ & ds=$90'$  & ds=$60'$ & ds=$45'$ \\
          & $A$, $p-$value & $A$, $p-$value & $A$, $p-$value & $A$, $p-$value \\
\hline
$l=[20,240]$ & 0.034(10), 0.000 & 0.035(10), 0.000 & 0.035(10), 0.000 & 0.036(10), 0.000\\
             & 0.037(10), 0.000 & 0.037(10), 0.000 & 0.038(10), 0.000 & 0.039(10), 0.000\\
             & 0.035(10), 0.000 & 0.035(10), 0.000 & 0.037(9), 0.000 & 0.037(9), 0.000\\
\hline
$l=[60,240]$ & - & 0.028(11), 0.007 & 0.028(11), 0.006 & 0.028(10), 0.003 \\
             & - & 0.028(10), 0.005 & 0.029(10), 0.001 & 0.029(10), 0.000 \\
             & - & 0.026(10), 0.006 & 0.028(10), 0.002 & 0.028(10), 0.001 \\
\hline
$l=[100,240]$ & - & 0.019(11), 0.173 & 0.018(11), 0.166 & 0.018(11), 0.165 \\
              & - & 0.017(11), 0.219 & 0.018(11), 0.168 & 0.019(11), 0.134 \\
              & - & 0.018(10), 0.148 & 0.019(10), 0.106 & 0.020(10), 0.086 \\
\hline
\end{tabular}
\caption{The amplitude of dipole, fit to normalized local variance maps, as obtained
         from PLANCK full mission \texttt{Commander} solution
         derived $E-$mode CMB map, are tabulated here. The first column lists
         the available multipole modes in the filtered CMB polarization maps
         as used in the present analysis, that
         correspond to probing HPA at different angular scales. Rest of the columns
         denote the four circular disc radii, as indicated, that are chosen to
         compute variances locally on the input map. The three sub-rows under any
         $l-$range denote the amplitude and its $p-$value as
         estimated using the three masks with $74\%$, $80\%$ and $85\%$ available
         sky fractions respectively. Note that the amplitude quoted here is `$A$'
         as referred to in Eq.~[\ref{eq:mod-temp-aniso}] and [\ref{eq:lve-dip-mod}].
         Quoted in braces are the error upto the last digits as indicated in the
         observed value of the dipole amplitude.}
\label{tab:dip-ampl-pval}
\end{table}

As can be readily noticed from Fig.~[\ref{fig:dip-mod-ampl}], the amplitude of dipole
in normalized LV maps from data denoted by triangle point types are significantly anomalous with
a probability, $p<1/1000$. So none of the 1000 normalized local variance maps from simulations
have a dipole of amplitude more than that seen in data. Further a careful observation also reveals
that when using smaller disc radii, the distance between data points and the footsteps of the
histogram on the right increases, in comparison to using disc radius of $120'$ (arcmin).
From Table~\ref{tab:dip-ampl-pval}, we also see that
the amplitudes are consistent with each other, rather independent of the galactic mask used
which have different masking fractions, as well as the disc radius chosen to compute
local variances.

\begin{figure}
\centering
\includegraphics[width=0.8\textwidth]{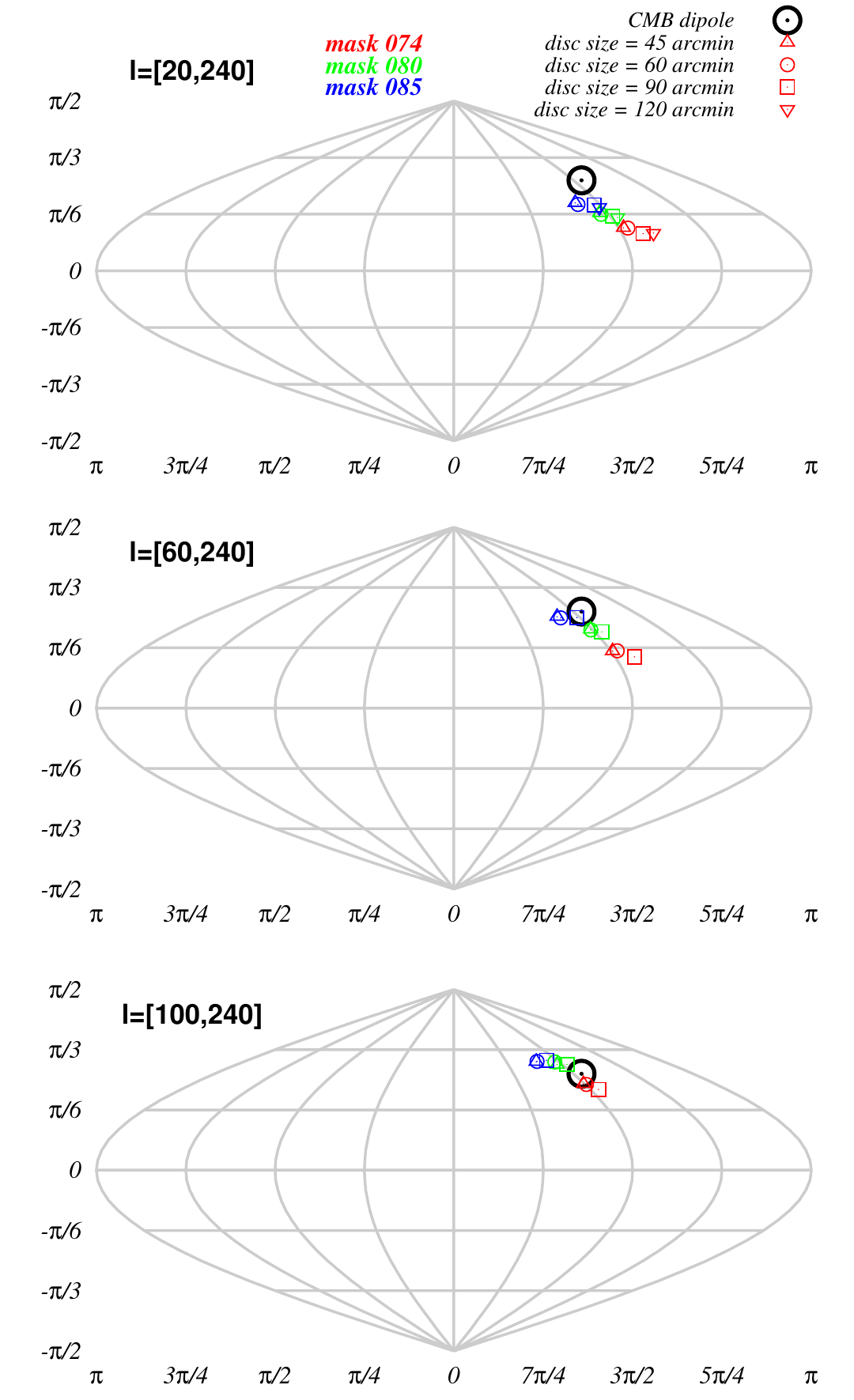}
\caption{Dipole directions inferred from normalized local variance maps of $E-$mode CMB map
         from PLANCK full mission observations, specifically obtained using the
         \texttt{Commander} component separation algorithm, are shown here in galactic co-ordinates.
         \emph{Top}, \emph{Middle} and \emph{Bottom}
         panels correspond to direction of dipoles as recovered from  three different
         filtered $E-$maps (ie., studying different angular scales)
         as indicated. The \emph{red}, \emph{green}, and \emph{blue} points
         in each panel denotes the dipole direction from LV maps as estimated using different
         masks with different sky fractions. The coloured \emph{up triangle}, \emph{circle},
         \emph{square} and \emph{inverted triangle}
         point types correspond to the four chosen circular disc radii, $45'$, $60'$, $90'$
         and $120'$ (arcmin), for computing local variances.}
\label{fig:dip-mod-direc}
\end{figure}

The direction of the dipoles as inferred from these normalized local variance maps
are shown in \emph{top panel} of Fig.~[\ref{fig:dip-mod-direc}]. They broadly point towards
the CMB kinetic dipole. The dipole directions inferred from data when using
galactic mask with larger sky fraction are aligned more closely towards CMB dipole.
This suggests that the normalized LV maps' dipole axes are sensitive to the available
sky fraction in the mask used, and consequently to the residual contamination.
still present in the CMB $E-$map.
Various LV maps' dipole directions ($\hat{d}_{LV}$) are listed in \emph{top} block of
Table~[\ref{tab:dip-mod-angl}]
along with the axes proximity to the CMB kinetic dipole
($\hat{d}_{CMB}=(\ell^\circ,b^\circ)=(264^\circ,48^\circ)$ \cite{kogut93})
direction measured as $\theta=\texttt{acos}(\hat{d}_{LV}\cdot\hat{d}_{CMB})$.

\begin{table}
\centering
\begin{tabular}{c c c c c}
\hline
$l-$range & ds=$120'$ & ds=$90'$  & ds=$60'$ & ds=$45'$ \\
          & $(\ell^\circ,b^\circ)$ $\theta^\circ$ & $(\ell^\circ,b^\circ)$ $\theta^\circ$ & $(\ell^\circ,b^\circ)$ $\theta^\circ$ & $(\ell^\circ,b^\circ)$ $\theta^\circ$ \\
\hline
$l=[20,240]$ & (253.0,20.0) 29.4 & (258.9,19.7) 28.6 & (265.1,22.6) 25.4 & (267.4,22.8) 25.4 \\
             & (266.5,28.3) 19.8 & (268.8,28.8) 19.5 & (274.5,30.0) 19.8 & (274.6,30.7) 19.1 \\
             & (271.9,33.7) 15.5 & (273.9,34.9) 15.1 & (283.6,35.1) 19.4 & (284.5,36.0) 19.2 \\
\hline
$l=[60,240]$ & - & (259.3,25.4) 22.8 & (266.5,28.5) 19.6 & (269.2,28.4) 20.0 \\
             & - & (265.6,37.9) 10.2 & (271.3,38.9) 10.5 & (271.1,39.2) 10.2 \\
             & - & (272.4,45.1)  6.5 & (284.1,44.9) 14.1 & (285.9,45.4) 15.2 \\
\hline
$l=[100,240]$ & - & (264.6,40.2)  7.8 & (269.3,42.6)  6.5 & (270.9,42.9)  7.0 \\
              & - & (266.2,52.6)  4.8 & (273.7,53.8)  8.4 & (274.7,52.5)  8.2 \\
              & - & (279.4,54.6) 11.6 & (288.4,54.1) 16.4 & (289.4,54.1) 17.0 \\
\hline
\end{tabular}
\caption{The dipole direction of LV maps obtained in the current study using $P-$masks, but utilizing
         various filtered $E-$maps and disc radii. Their amplitudes are listed in Table~[\ref{tab:dip-ampl-pval}].
         The three sub-rows in each horizontal block correspond to using the masks \texttt{mask074},
         \texttt{mask080} and \texttt{mask085} respectively.
         $(\ell,b)$ (in degrees) correspond to the dipole directions in galactic coordinates, and
         $\theta=\texttt{acos}(\hat{d}_{LV}\cdot\hat{d}_{CMB})$, where $\hat{d}_{LV}$ is the dipole direction
         of a local variance map and $\hat{d}_{CMB}$ is the CMB kinetic dipole direction. $\theta$ denotes
         an LV dipole axis's proximity to the CMB kinetic dipole. The angles are given in degrees ($^\circ$).}
\label{tab:dip-mod-angl}
\end{table}

\begin{figure}
\centering
\includegraphics[width=0.8\textwidth]{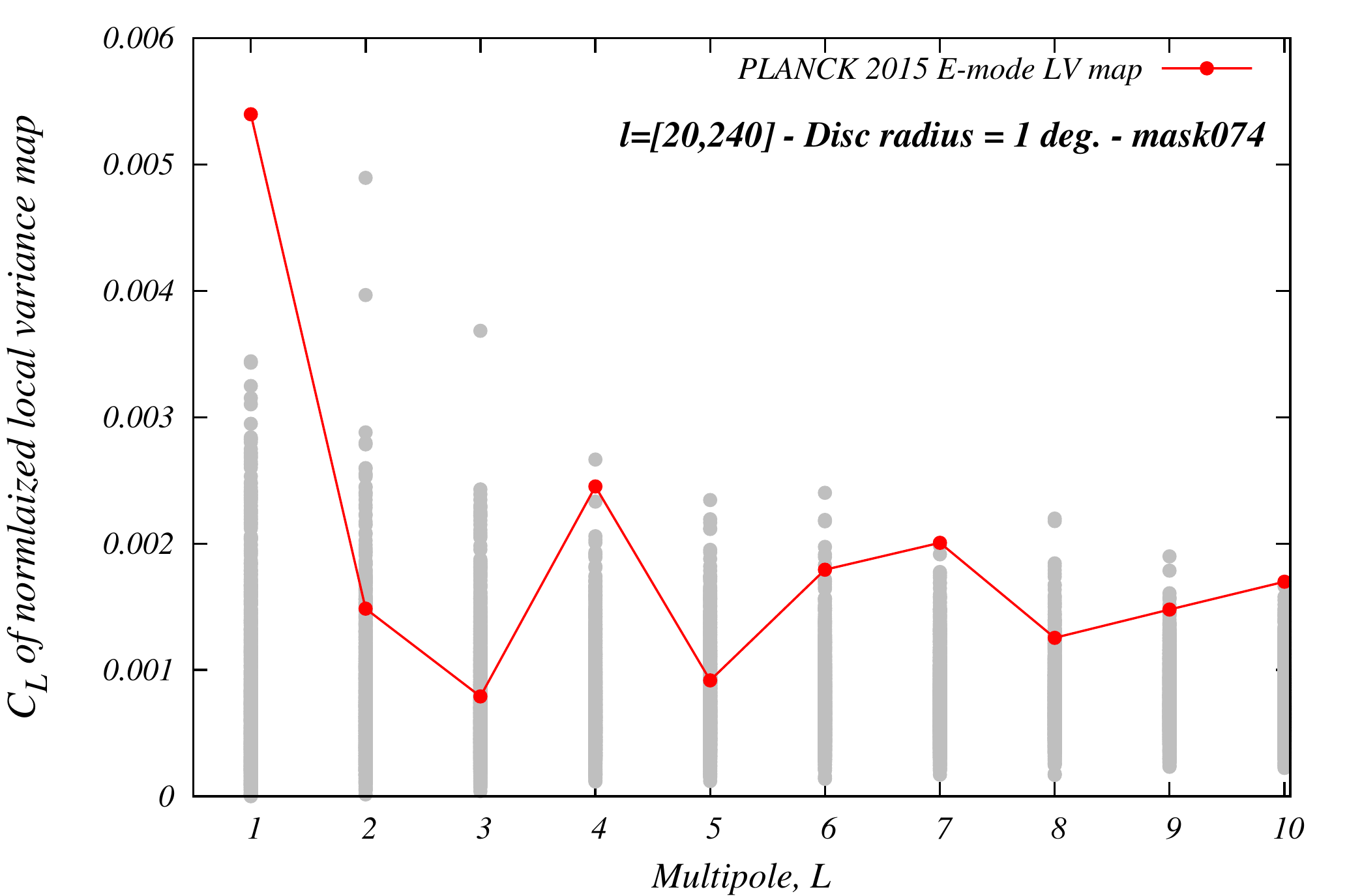}
\caption{Angular power spectrum of local variance map obtained using $E-$mode
         CMB map from PLANCK's \texttt{Commander} 2015 CMB estimate, comprising the multipole
         $l=[20,240]$. Specifically a circular disc of $60'$ (arcmin) radius and
         \texttt{mask074} were used. We see a significant power in only the $L=1$ mode.
         This example plot illustrates that the power asymmetry signal seen in polarization
         data is entirely dipolar in nature. The grey dots against each multipole, $L$,
         denote simulation estimates, and the red line with red point type denotes angular power
         spectrum of LV map from data as indicated.}
\label{fig:lv-map-cl}
\end{figure}

Now we check whether there is any significant power in modes other than dipole in the local
variance maps. This would inform us whether the anomalous power asymmetry we found is only
dipolar in nature or other modes are also contributing to the observed power asymmetry.
We find that there is significant power only in the dipole component of the local variance
maps from polarization data.
As an example, in Fig.~[\ref{fig:lv-map-cl}], we show the angular power spectrum
plot of local variance map derived using the range $l=[20,240]$ from data, and estimated
using disc radius of $60'$ (arcmin) and the galactic mask \texttt{mask074}.
We applied the first order $f_{sky}$ correction to estimate the
angular power spectrum, $C_L$, where $L$ is the multipole index of LV maps.

Thus we find a signal of anomalous dipole power asymmetry in CMB $E-$mode polarization maps
from PLANCK full mission data on large angular scales, broadly pointing in the
CMB kinetic dipole direction. The dipole axis inferred is however found to be sensitive to
the galactic mask, used to estimate the LV maps from the current release of polarization data.

\subsection{Scale dependence of amplitude of dipolar power asymmetry}

In this section we probe the dipole power asymmetry signal at small angular scales.
This will allow us to evaluate any scale dependence in the amplitude of
the dipole modulation signal we are studying in CMB $E-$mode polarization maps.
Given the limited number of circular disc radii we could use with the current data,
we exclude the large angular scales in progression to study any such scale
dependence with local variance maps.

To this end, we use two filtered CMB $E-$mode maps containing small angular scales
ie., comprising the multipoles $l=[60,240]$ and $l=[100,240]$. A high and low pass
filters defined in Eq.~[\ref{eq:highpassfilter}] and [\ref{eq:lowpassfilter}]
are applied over a width of 20 multipoles at both ends of the multipole
ranges mentioned for the two filtered maps.
Since these maps contain smaller angular scales, we use only three circular
discs of radii $90'$, $60'$ and $45'$ (arcmin) to get LV maps. The
three $P-$masks shown in Fig.~[\ref{fig:masks}] are used as in the previous section.
We again demand availability of atleast $90\%$ of pixels in the effective
circular discs, defined as union of the galactic masks and the full circular
discs defined locally on \texttt{HEALPix} $N_{side}=64$ grid to map local
variances.

\begin{figure}
\centering
\includegraphics[width=0.95\textwidth]{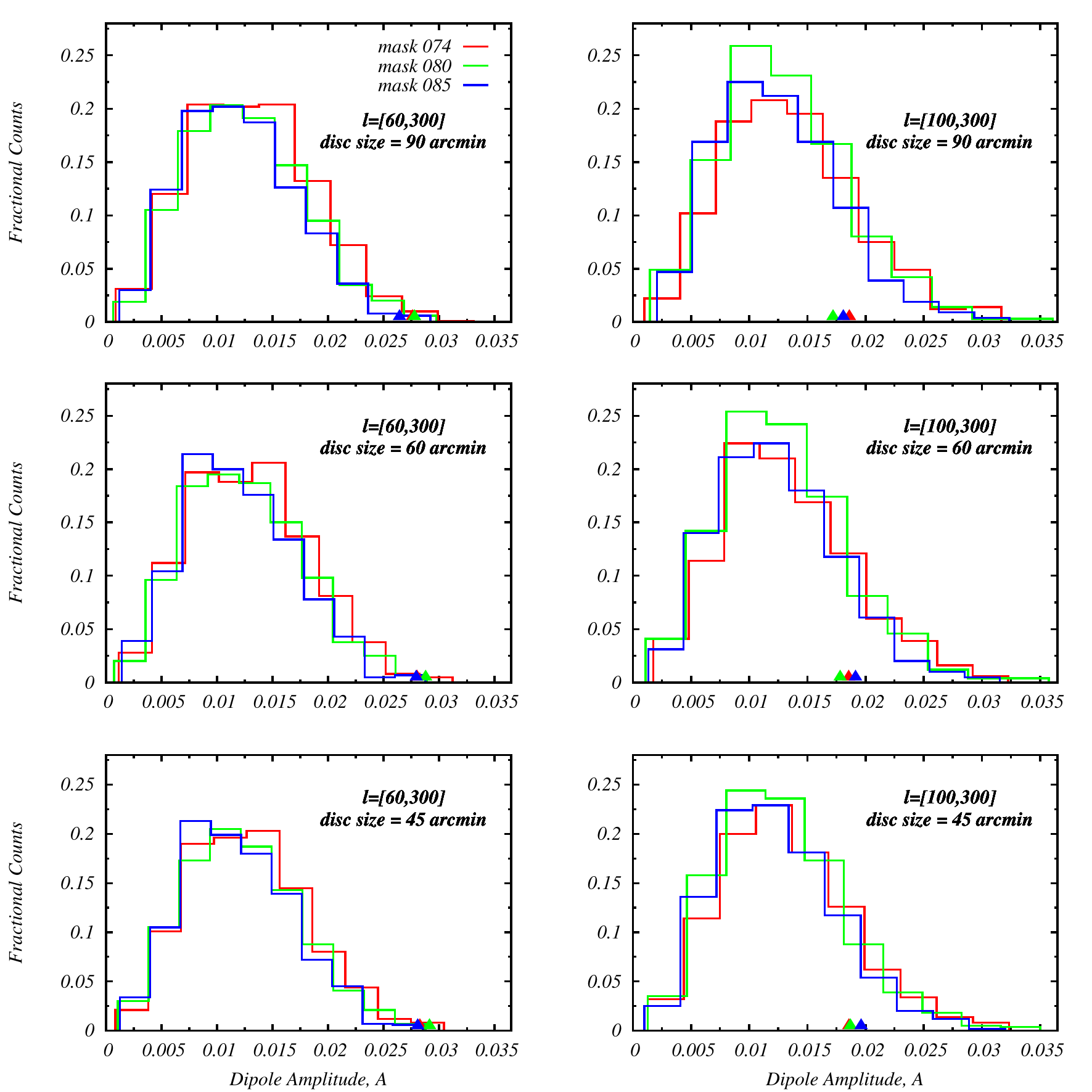}
\caption{Same as Fig.~[\ref{fig:dip-mod-ampl}], but for normalized local variance maps
         obtained using the mutlipole range $l=[60,240]$ and $l=[100,240]$ (the two columns),
         and local circular disc radii of radii $90'$, $60'$ and $45'$ (arcmin) (the three rows).
         Each plot has three histograms that correspond to estimating the dipole
         amplitudes with three different $P-$masks as labelled.}
\label{fig:dip-mod-ampl-l60-l90}
\end{figure}

The amplitude of dipole of normalized local variances maps from the
two filtered CMB $E-$mode maps are shown in Fig.~[\ref{fig:dip-mod-ampl-l60-l90}].
The plots in left-hand column of the figure corresponds to using the filtered map containing
$l=[60,240]$, and the right-hand column corresponds to those comprising $l=[100,240]$.
Each plot has again three histograms in \emph{red}, \emph{green} and \emph{blue}
corresponding to use of \texttt{mask074}, \texttt{mask080} and \texttt{mask085}
galactic masks. The data values are denoted by triangle point types in each plot.
The three rows in Fig.~[\ref{fig:dip-mod-ampl-l60-l90}] correspond to results obtained
using the three disc radii $90'$, $60'$ and $45'$ (arcmin).

One can see that, the amplitude of dipole in LV maps is anomalous only when obtained from the
range $l=[60,240]$. The dipole amplitudes in the LV maps from the filtered $E-$map
containing $l=[100,240]$ modes are consistent with amplitude excursions expected in
isotropic maps. The dipole power asymmetry amplitudes and their $p-$values corresponding to
the two filtered maps are listed in the \emph{second} and \emph{third} block of
Table~[\ref{tab:dip-ampl-pval}].
The corresponding dipole directions are shown in the \emph{middle} and \emph{bottom}
panel of Fig.~[\ref{fig:dip-mod-direc}].
These dipole directions are still broadly pointing
towards CMB Doppler boost direction, nearly independent of the choice of disc radius
and galactic mask used. But the orientation of the dipole axes move towards higher latitudes
around the kinetic dipole axis by excluding low multipoles.
The dipole directions corresponding to LV maps from these two filtered CMB $E-$mode maps are
listed in the \emph{second} and \emph{third} block of Table~[\ref{tab:dip-mod-angl}], along
with their angular separation from the CMB kinetic dipole direction.

\subsection{LV analysis using $E-$mode mask}
\label{apdx:emask-anls}
Here we redo the exercise of local variance analysis described in the preceding sections,
using $E-$mode masks. So far we used masks derived in $Q/U$ basis ie. $P-$masks,
where $P^2=Q^2 + U^2$.
Now we repeat the LV analysis using $E-$mode foreground masks created as described below,
as a test of consistency of our results obtained using $P-$maps.

\begin{figure}
\centering
  \includegraphics[width=0.75\textwidth]{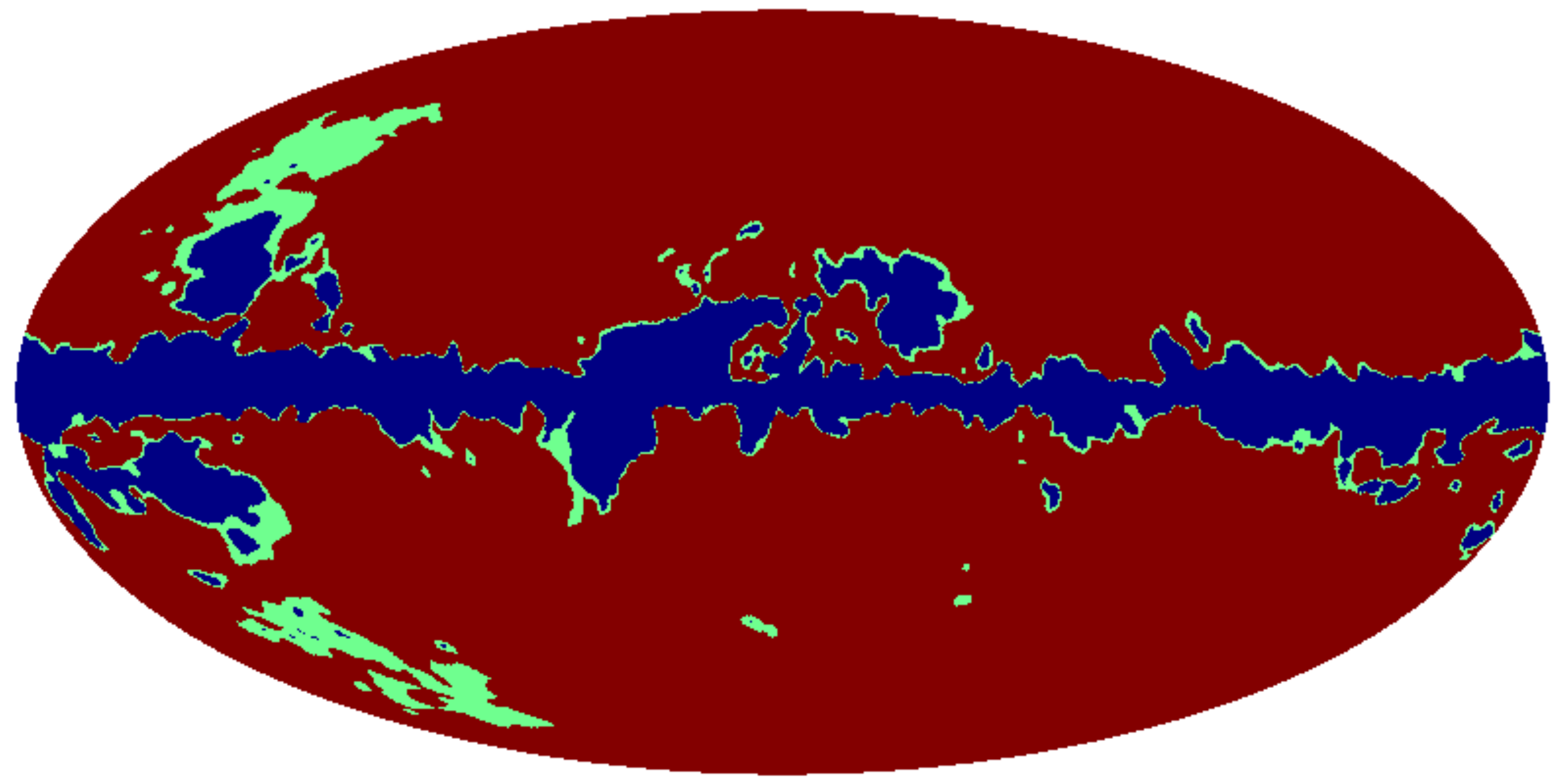}
\caption{The two $E-$mode masks viz., \texttt{mask080E} (red only)
         and \texttt{mask085E} (light green+red) are shown here. The blue region is
         commonly excluded by both masks.}
\label{fig:emasks}
\end{figure}

Once we obtain the difference maps \texttt{030-070} and \texttt{353-070} from
raw PLANCK stokes $Q/U$ satellite maps, as described in Sec~[\ref{sec:pmasks}],
we convert these difference maps into $E-$mode maps which contain only foregrounds.
The two $E-$mode difference maps are combined to form a \emph{junk map}, where
we retain only the largest absolute pixel-value, among the two, in the output
\emph{junk map}, following the procedure of Tegmark et al. (2003) \cite{tegmark03}.
This map is then thresholded such that the resultant masks contains $f_{sky}\approx 80\%$
and $85\%$ ie., unmasked sky fractions. The masks thus obtained, denoted here after as
\texttt{mask080E} and \texttt{mask085E}, are shown in Fig.~[\ref{fig:emasks}].

We now go on to estimate the local variance maps using these $E-$mode masks.
The analysis procedure is same as described in Sec.~[\ref{subsec:lvelowl}], except
for using the newly derived $E-$masks. Here we only analyze the filtered
$E-$mode CMB map that includes the large angular scales ie., the data $E-$maps containing
the full range $l=[20,240]$, and also use a single disc radius of $120'$ (arcmin)
to compute local variances.

\begin{figure}
\centering
  \includegraphics[width=0.48\textwidth]{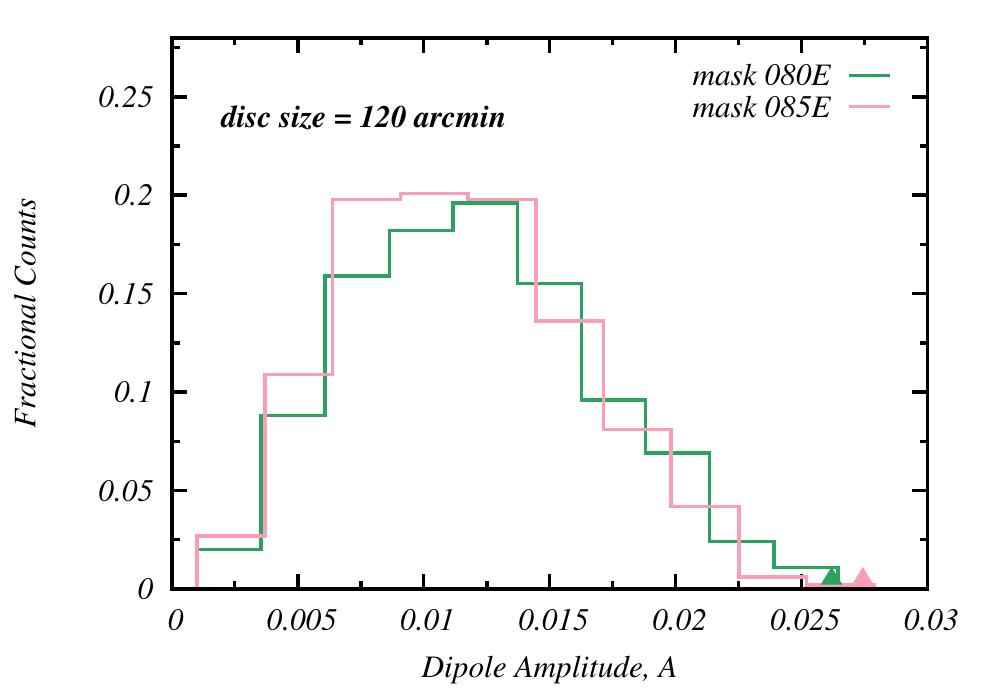}
  ~
  \includegraphics[width=0.48\textwidth]{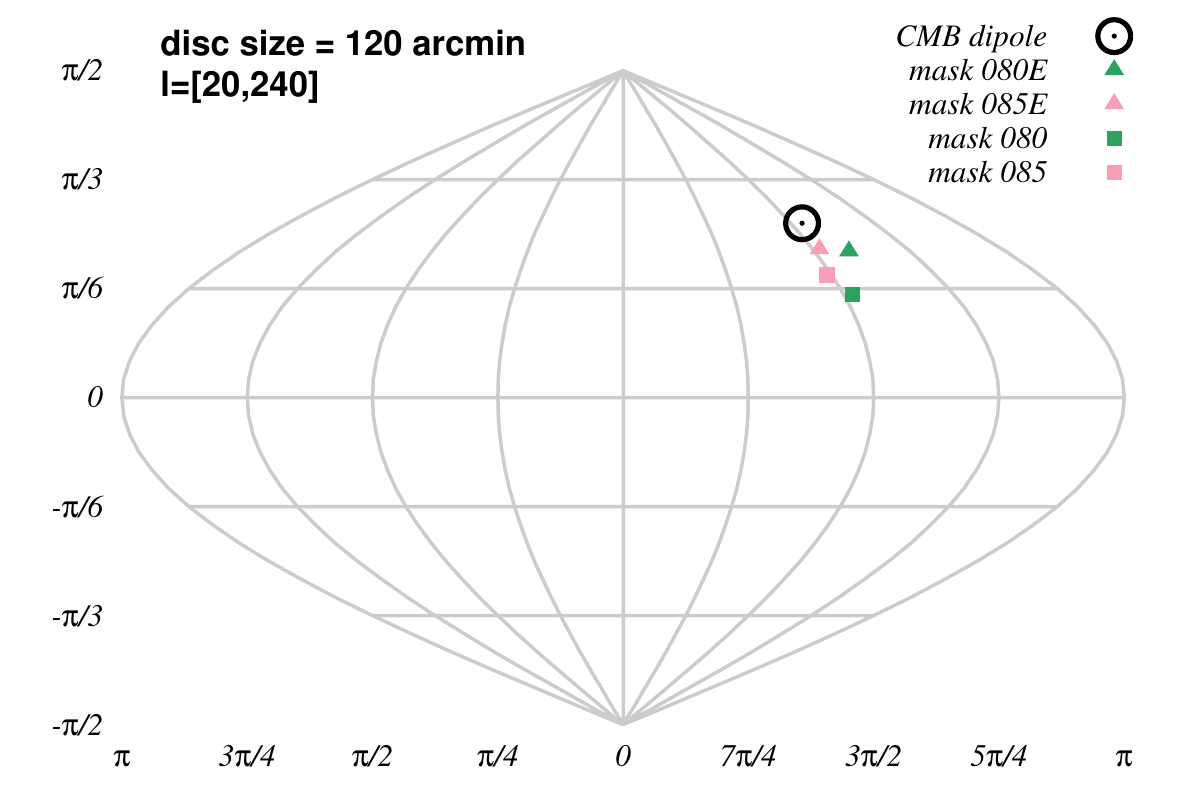}
\caption{\emph{Left} : Histograms of dipole amplitudes obtained from simulations
         in comparison to data derived values, when using \texttt{mask080E} and \texttt{mask085E}
         are shown here.
         \emph{Right} : Dipole directions inferred from normalised LVE maps obtained
         using the two $E-$space masks and the filtered $E-$mode data CMB map containing
         the large angular scales ($l=[20,240]$) are shown here.
         These LV maps are estimated using $120'$ (arcmin) radius circular discs, to compute
         local variances.
         For comparison, the dipole directions shown in the top panel of Fig.~[\ref{fig:dip-mod-direc}]
         corresponding to LVE maps obtained using CMB $E-$map with $l=[20,240]$,
         and employing the $P-$masks viz., \texttt{mask080} and \texttt{mask085}
         and circular radius of $120'$ (arcmin) are also plotted.}         
         \label{fig:ampl-angl-emask}
\end{figure}

The dipole amplitudes derived by fitting a dipole to normalized local variance maps
in this case are shown in \emph{left-hand} panel of Fig.~[\ref{fig:ampl-angl-emask}]
in comparison to simulations.
The two histograms in that panel correspond to the two $E-$masks used, and the
data points are denoted by triangle point types.
The direction of the dipole as inferred from these normalized data LVE maps
are shown in the same figure in the \emph{right-hand} panel.
The observed dipole amplitudes and their significances using \texttt{mask080E}
and \texttt{mask085E} are tabulated in Table~\ref{tab:dip-ampl-pval-emask}.

The dipole amplitudes now obtained using $E-$space masks are found to be lower
in amplitude than those obtained when using $P-$masks in Sec.~[\ref{subsec:lvelowl}].
Their significances have become $0.3\%$ and $0.1\%$, respectively, when
\texttt{mask080E} and \texttt{mask085E} are used. The dipole directions now seem to be
more closely aligned with the CMB kinetic dipole direction. For reference, the dipole
directions from using the $P-$masks viz., \texttt{mask080} and \texttt{mask085} with
the same circular radius of $120'$ (arcmin) to obtain LV maps are also
indicated in the right panel of Fig.~[\ref{fig:ampl-angl-emask}].

\begin{table}
\centering
\begin{tabular}{r c c c c}
\hline
mask & $A$ & $p-$value & $\hat{d}_{LV}=(\ell^\circ,b^\circ)$ & $\theta^\circ=\texttt{acos}(\hat{d}_{LV}\cdot\hat{d}_{CMB})$ \\
\hline
\texttt{mask080E} & 0.026(10) & 0.003 & (253.7,40.2) & 10.7 \\
\texttt{mask085E} & 0.027(9) & 0.001 & (266.8,40.8) & 7.5 \\
\hline
\end{tabular}
\caption{Dipole amplitude ($A$) and significances ($p$) derived from LVE maps obtained using
         PLANCK \texttt{Commander} $E-$mode CMB map, and the \texttt{mask080E}
         and \texttt{mask085E} confidence masks, are tabulated here. The CMB
         $E-$map contains the multipoles $l=[20,240]$ and a local circular
         discs of radius $120'$ (arcmin) were used to compute variances.
         Also given are the direction of LV dipoles ($\hat{d}_{LV}$) and their          
         angular distance from CMB Doppler direction, both in degrees.}
\label{tab:dip-ampl-pval-emask}
\end{table}

\section{Conclusions}
In this paper, we tried to probe the presence of hemispherical power asymmetry
in CMB $E-$mode polarization data, that was so far studied in CMB temperature data.
To this end we used Stokes $Q/U$ CMB maps from PLANCK full mission data estimated
using \texttt{Commander} component separation procedure. From these full-sky Stokes
$Q/U$ maps available at \texttt{HEALPix} $N_{side}=1024$ and having a Gaussian beam
smoothing of $FWHM=10'$ (arcmin), we derived $E-$mode polarization map of CMB. This
$E-$mode map is obtained at a low resolution of $N_{side}=256$ and with a smoothing
given by a Gaussian beam of $FWHM=40'$ (arcmin). The low resolution $E-$mode map is further
low-pass and high-pass filtered to get three maps comprising different ranges of multipoles
viz., $l=[20,240]$, $[60,240]$ and $[100,240]$ to test for any scale dependence
of the amplitude of hemispherical power asymmetry modelled as dipole modulation.

We find  a tentative evidence for the presence of dipole modulation in
polarization data as well, with an amplitude of $\sim 2.6\% - 3.9\%$ depending
on the mask, and disc radius used on large angular scales (ie., using the multipole range
$l=[20,240]$). The lower values are found when $E-$space masks are used for
computing local variance maps, instead of $P-$masks where $P^2=Q^2+U^2$.
This sheds light on some important analysis aspect that the mask used for
analyzing CMB $E-$mode maps have to be constructed carefully, and the $P-$masks
that are provided with second PLANCK data release may not be readily applicable
in $E-$space.
LV maps from intermediate and small angular scales of the CMB $E$ polarization map
considered in the present work are found to have lower dipole amplitude.
The amplitude of the dipoles on large angular scales were found to be very
significant with a probability of $p<1/1000$, which goes down to $p\sim 0.2$  on small
angular scales.
The direction of the dipoles inferred from fitting a dipole to the local variance maps
are found to be roughly pointing towards the CMB kinetic dipole direction, and
broadly independent of the angular scales involved, and masks and circular disc
radii used to obtain LV maps. The dipole axis inferred when using $E-$space masks
are found to be more closely aligned with the CMB dipole direction.

However, with the know issue of systematic artefacts on large angular scales of PLANCK
polarization data,
only precise future measurements of polarization or a future PLANCK plarization data release
with carefully processed large angular scales will validate these findings, beyond any systematics
still present in the multipole range studied here.

\section*{Acknowledgements}
We are very thankful to Anthony Banday for a careful reading of our draft and giving his valuable comments. We also thank Changbom Park for helpful discussions.
P.K.A. and A.S. would like to acknowledge the support of the National Research Foundation of Korea (NRF-2016R1C1B2016478).
The present work has extensively made use of the publicly available \texttt{HEALPix}\footnote{\url{https://healpix.jpl.nasa.gov/}} \cite{healpix} package, a frame work for representing data on a sphere and to perform additional manipulations thereof.
Part of the results presented here are based on observations obtained with
PLANCK\footnote{\url{http://www.esa.int/Planck}}, an ESA science mission with instruments
and contributions directly funded by ESA Member States, NASA, and Canada.



\appendix
\section*{\LARGE Appendix}

\section{Treatment of noise simulations}
\label{apdx:noisesim}
As described in Sec.~[\ref{sec:sim}], effective noise realizations were obtained using
co-addition of smoothed, downgraded frequency specific FFP8  noise maps. However the
noise levels in polarized FFP realizations were under estimated compared to data.
One can use the half-mission half-difference (HMHD) map from the two \texttt{Commander}
polarized CMB maps derived using half mission data as proxy
to the true noise level in PLANCK \texttt{Commnader} $E-$mode CMB map.
These half-mission maps are also provided through PLANCK second public release\footnote{\url{http://irsa.ipac.caltech.edu/data/Planck/release_2/all-sky-maps/matrix_cmb.html}}.
This HMDH map is processed in the same way as the data CMB $E-$mode polarization map,
to get a low resolution version.

\begin{figure}
\centering
\includegraphics[width=0.75\textwidth]{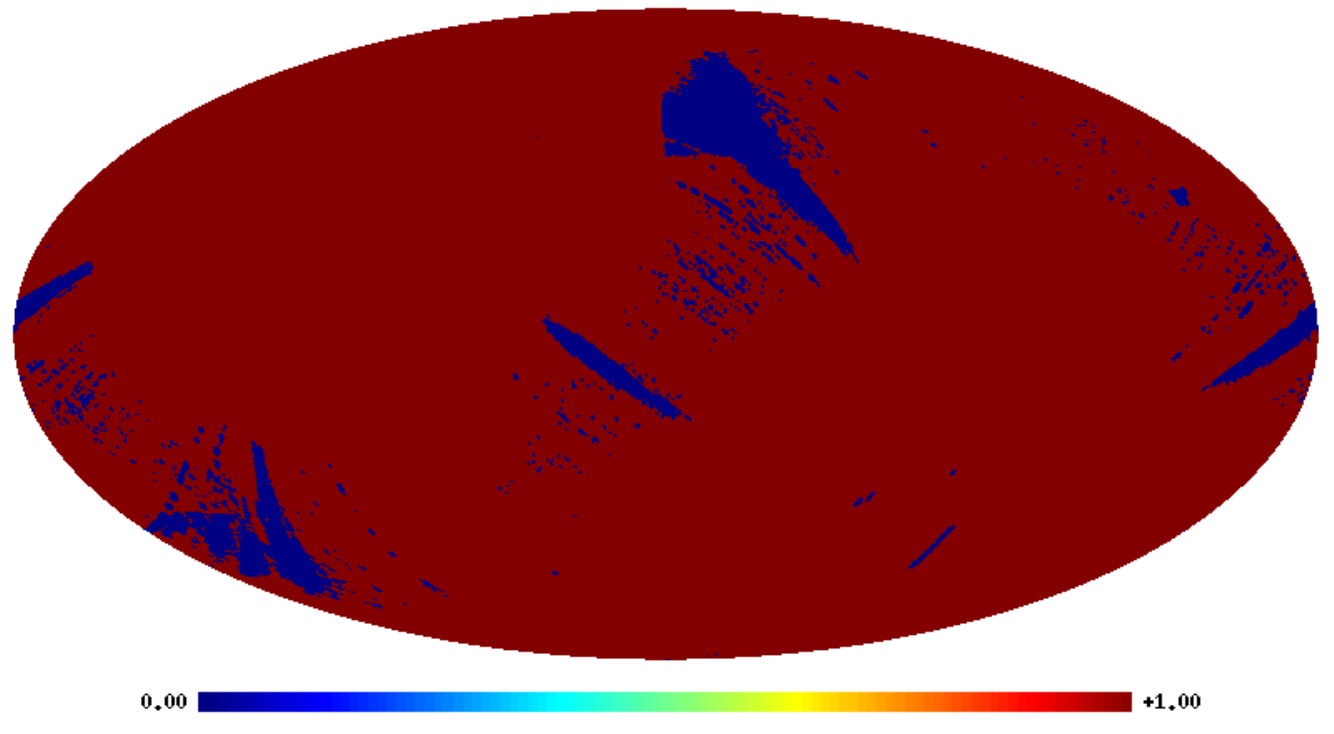}
\caption{Half-mission mask that is suitably extended to $N_{side}=256$ is shown here. The original
         mask is provided at \texttt{HEALPix} $N_{side}=1024$. See text for details.}
\label{fig:hmmask}
\end{figure}

Here we compare the noise levels in simulations and data, in an attempt to be able
to use the FFP polarization noise simulations with reasonable modifications. To reliably derive
the noise angular power spectrum from \texttt{Commnader} polarized HMHD maps, and from the
FFP8 polarized noise simulations, we have to first modify the masks shown in Fig.~[\ref{fig:masks}].
The half-mission (HM) data itself comes with additional mask owing to missing pixels in respective
data sets, that is further processed and downgraded to $N_{side}=256$. The original half-mission
mask is made available at $N_{side}=1024$.
To this we also add the missing pixels mask from single year maps to make it more conservative.
This combined mask is downgraded to $N_{side}=256$ and smoothed
with a Gaussian beam of $FWHM=40'$ (arcmin). Then a pixel cut-off of 0.95 is imposed whereby
pixels with pixel-value greater than this are set to `$1$' and rest to `$0$'. The
mask thus obtained is shown in Fig.~[\ref{fig:hmmask}].

The three $P-$masks shown in Fig.~[\ref{fig:masks}] are then taken in \emph{union} with
the suitably extended, downgraded half-mission mask of Fig.~[\ref{fig:hmmask}]. However in order
to compute the noise power spectrum from partial sky, it would be useful to apodize the masks to
avoid ringing in the recovered noise power spectra due to sharp mask boundaries (the so-called Gibbs phenomenon). So we use a \emph{cosine}
profile to apodize the three \emph{union} (galactic+HM) masks over a width of $200'$ (arcmin) at the
mask boundaries. Recall that these masks are at $N_{side}=256$, which have pixels of
side $\sqrt{{4\pi}/(12\times 256^2)} \approx 13.74'$ (arcmin), assuming them to be square shaped.
So, choosing to apodize over a width of 10 pixels strips along the mask boundary equals
an apodization width of $\sqrt{2}\times13.74'\times 10 \approx 194.3'$.

\begin{figure}
\centering
\includegraphics[width=0.91\textwidth]{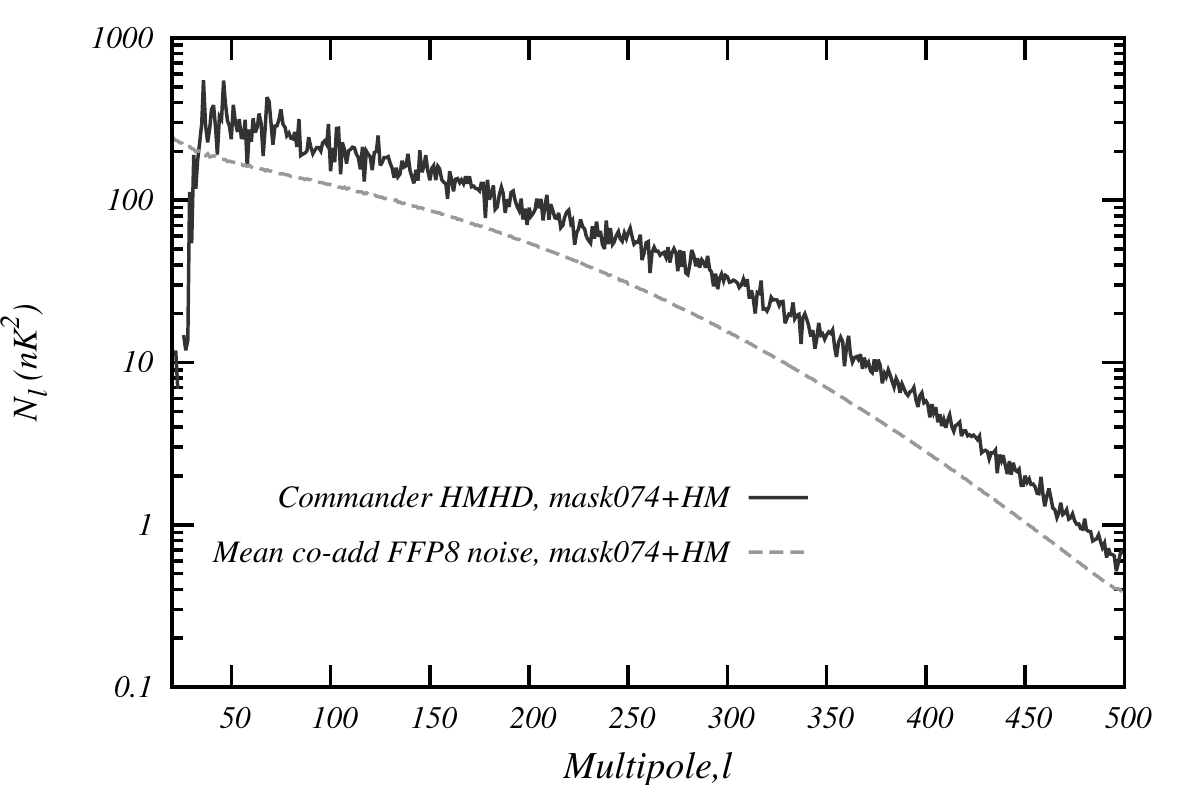}
~
\includegraphics[width=0.9\textwidth]{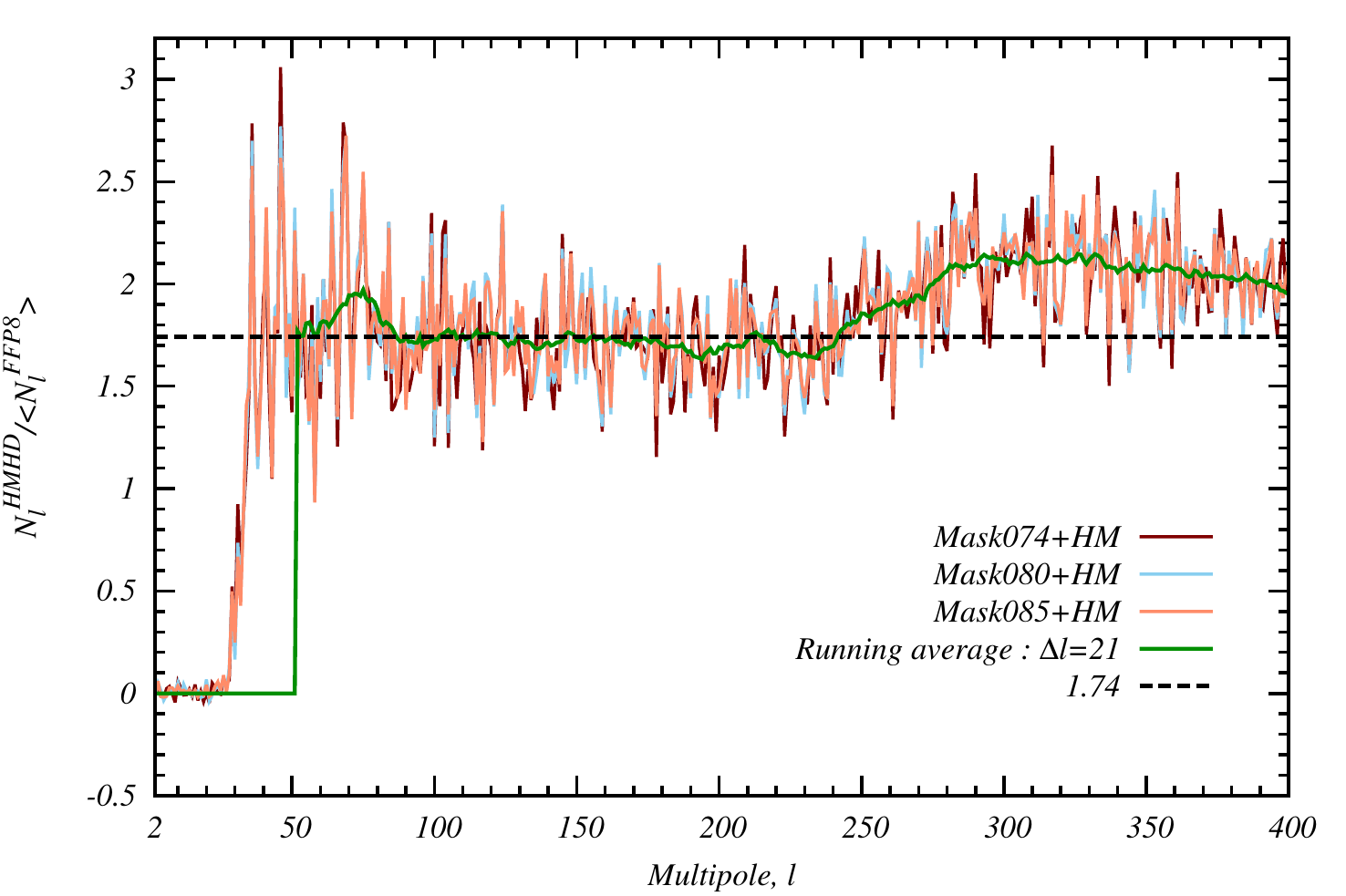}
\caption{Comparison of noise levels in polarization data and FFP8 simulations are shown here,
         obtained using the three galactic masks shown in Fig.~[\ref{fig:masks}], taken
         in union with the half-mission mask of Fig.~[\ref{fig:hmmask}]. In practice, apodized versions
         of the said masks were used to compute the noise power spectra.}
\label{fig:nl-data-sim-074HM}
\end{figure}

These apodized union masks are used to mask the \texttt{Commander} HMHD
$E-$mode map. Then the pseudo full-sky noise power spectra were obtained following
the MASTER algorithm~\cite{hivon02}. We repeat the same exercise on 1000 FFP8 co-added
noise polarization maps by applying the union masks with differing sky fractions.
A comparison of respective power spectra is shown in \emph{top panel} of
Fig.~[\ref{fig:nl-data-sim-074HM}].
For brevity only the pseudo full-sky noise power spectrum from using the union mask
\texttt{mask074+HM} on data and simulations is shown in that panel.

In the \emph{bottom panel} of Fig.~[\ref{fig:nl-data-sim-074HM}], we show the ratio
of pseudo full-sky noise power spectrum from \texttt{Commander} HMHD $E-$mode map ($N_l^{HMHD}$)
to the corresponding mean pseudo full-sky noise spectrum from 1000 effective FFP noise
maps ($\langle N_l^{FFP8}\rangle$), obtained using the three union masks. Also shown
is the mean of the three noise power spectra ratios (as dark green solid line), smoothed using
a running window of $\Delta l = 21$ (ie., average over $l\pm10$ at each multipole, $l$).
We find that this mean of the ratio of noise power spectra is approximately
constant upto $l=240$. By taking the average of all three noise power spectra ratios in
the range $l=[41,240]$ we get a factor $\approx 1.74$, that is plotted as a black dashed
curve in the same figure (bottom panel). Thus we scale the effective noise maps by a factor of
$\sqrt{1.74}\approx1.32$ to alleviate their noise levels in comparison to true noise in data
maps.

As explained in Sec.~\ref{apdx:emask-anls}, we devised additional masks directly
in $E-$space. Since the $P-$masks we used have varying mask fractions, we expect
the scaling factor found with them to still hold when estimated using $E-$masks.
Nevertheless, we analyzed the HMHD noise proxy using \texttt{mask085E} (see Fig.~[\ref{fig:emasks}])
following the same procedure as with $P-$masks. We find the ratio of the \texttt{Commander} HMHD
to mean simulation noise power spectra obtained by employing \texttt{mask085E} to have a similar
level of mismatch, and also result in approximately same factor of $1.32$ for scaling
the simulations.

\end{document}